\documentclass[aps,prb,showpacs,twocolumn,superscriptaddress]{revtex4}
\usepackage{amsmath}
\usepackage{amsfonts}
\usepackage{amssymb}
\usepackage{graphicx}
\begin{document}
  \title{Classical and quantum two-dimensional anisotropic Heisenberg antiferromagnets}
  \author{M.~Holtschneider}
  \affiliation{Institut f\"ur Theoretische Physik,
    RWTH Aachen,
    52056 Aachen, Germany}
  \author{S.~Wessel}
  \affiliation{Institut f\"ur Theoretische Physik III,
    Universit\"at Stuttgart,
    70550 Stuttgart, Germany}
  \author{W.~Selke}
  \affiliation{Institut f\"ur Theoretische Physik,
    RWTH Aachen,
    52056 Aachen, Germany}
  \begin{abstract}
    The classical and the quantum, spin~$S=\frac{1}{2}$, versions of the 
    uniaxially anisotropic Heisenberg antiferromagnet on a square lattice in a 
    field parallel to the easy axis are studied using Monte Carlo techniques. 
    For the classical version, attention is drawn to biconical
    structures and fluctuations at low 
    temperatures in the transition region between the antiferromagnetic and 
    spin-flop phases. For the quantum version, the previously 
    proposed scenario of a first-order transition between the 
    antiferromagnetic and spin-flop phases with a critical endpoint and a 
    tricritical point is scrutinized.    
  \end{abstract}
  \maketitle
\section{Introduction}
\label{sec_in}

Uniaxially anisotropic Heisenberg antiferromagnets in an external field along 
the easy axis have attracted much interest, both theoretically and 
experimentally, due to their interesting structural and critical properties. 
In particular, they display a spin-flop phase, and multicritical behavior 
occurs at the triple point of the antiferromagnetic (AF), spin-flop (SF) and 
paramagnetic phases.\cite{fnk,ro,lb,tk,gj,cab,kst,chl,ou,cu}

A prototypical model for such antiferromagnets is the XXZ model, with the 
Hamiltonian
\begin{equation}
  \mathcal{H} \; = \; 
  J \sum\limits_{(i,j)}\left[ \, \Delta (S_i^x S_j^x + S_i^y S_j^y) + S_i^z S_j^z \, \right] \; - \; 
  H \sum\limits_{i} S_i^z \quad \text{,}
  \label{eq_ham}
\end{equation}
where the sum runs over neighboring spins of a cubic, dimension~$d=3$, or 
square lattice, $d=2$. The coupling constant~$J$ and the field~$H$ are 
positive; the anisotropy parameter~$\Delta$ may range from zero to one. 
Furthermore, 
$S_i^x$, $S_i^y$, and~$S_i^z$ denote the spin components at lattice site~$i$.

For the three-dimensional case, early renormalization group 
arguments\cite{fnk} and Monte Carlo simulations\cite{lb} suggested that the 
triple point is a bicritical point with $O(3)$~symmetry. Only a few years ago, 
this scenario has been questioned, based on high-order perturbative 
renormalization group calculations.\cite{cbv} It has been predicted that there 
may be either a first order transition, or that the 'tetracritical 
biconical' \cite{fnk} fixed point, due to an intervening 'mixed' or
'biconical' phase in between the AF and SF phases \cite{gor,pt,lf}, may
be stable.

In two dimensions, conflicting predictions on the nature of the triple point 
have been put forward recently \cite{ho,zl,pv,st}, when analyzing the
classical version of the 
above model with spin vectors of unit length, and the quantum version with 
spin~$S=\frac{1}{2}$.

Indeed, in the classical case, simulational evidence for a
narrow (disordered) phase between the AF and SF phases has
been presented \cite{ho}, extending presumably
down to zero temperature. \cite{zl} On the other hand, in
the quantum case, based on simulations as well, a direct transition
of first order between the AF and SF phases has been argued
to occur at low temperatures. \cite{st,koh,yun}

Obviously, experimental data have to be viewed with care because deviations 
from the XXZ Hamiltonian, Eq.~(\ref{eq_ham}), such as crystal field 
anisotropies or longer-range interactions, may affect relevantly the critical 
behavior of the triple point.\cite{gj,lf,lsk,gor,pt,ba}

In the following, we present results from large-scale Monte Carlo 
simulations of the XXZ model on a square lattice for both the classical and 
the quantum variant. In the quantum Monte Carlo simulations, the method of 
the stochastic series expansion (SSE)\cite{sk} is used, and the standard 
Metropolis algorithm is applied for the classical case. The simulations are 
augmented by a ground-state analysis of the classical model, showing the 
significance of biconical structures. The outline of the paper is as follows: 
First we shall discuss our findings on the classical model, followed by a 
section on the quantum version of the XXZ model. A summary concludes the paper.

\section{Classical model}
\label{sec_cl}
\begin{figure}
  \includegraphics{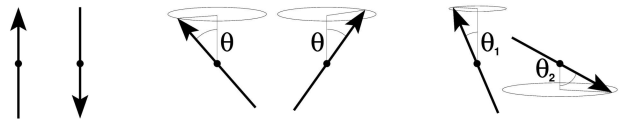}
  \caption{Ground state configurations of the classical model sketched by 
    the directions of spins on the two sublattices (i.\,e. at neighboring 
    sites), from left to right: AF, SF, and biconical state. The circles 
    denote the trivial degeneracy in the $xy$-plane.}
  \label{fig_gstate}
\end{figure}

The ground states of the classical model on a square lattice, see 
Hamiltonian~(\ref{eq_ham}), can be determined exactly. The AF structure is 
stable for magnetic fields below the critical value
\begin{equation}
  H_{\text{c}1} \; = \; 4 J \sqrt{1-\Delta^2} \quad \text{,}
  \label{eq_hc1}
\end{equation}
while for larger fields the SF state is energetically favorable. 
At~ $H_{\text{c1}}$, the tilt angle~$\theta_{\text{SF}}$ of the SF structures, 
see Fig.~\ref{fig_gstate}, is given by
\begin{equation}
  \theta_{\text{SF}} \; = \; \arccos \sqrt{\frac{1-\Delta}{1+\Delta}} \quad \text{.}
  \label{eq_sfangle}
\end{equation}
Increasing the field beyond \mbox{$H_{\text{c}2} = 4J(1+\Delta)$}, all spins 
perfectly align in the $z$-direction.
\begin{figure}
  \includegraphics{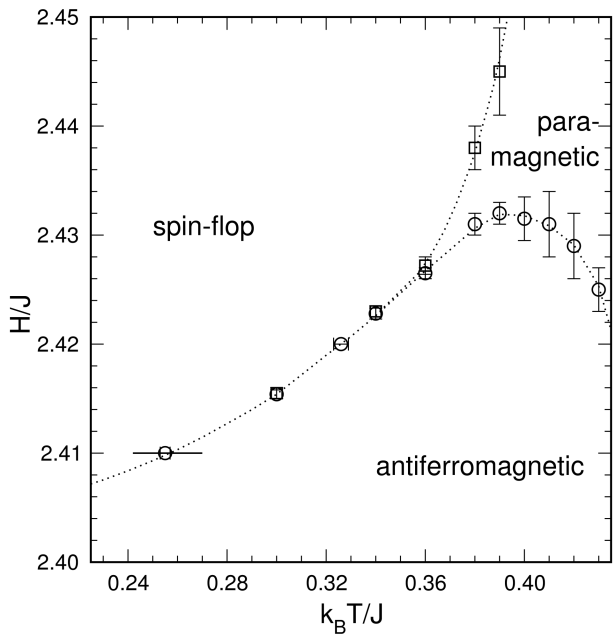}
  \caption{Detail of the phase diagram of the XXZ model on a square lattice 
    with~$\Delta=\frac{4}{5}$, see Ref.~\onlinecite{ho}. Squares refer to the 
    boundary of the SF, circles to that of the AF phase. The solid line refers 
    to the magnetic field~$H/J=2.41$, where the probability 
    distribution~\mbox{$P(\theta_m,\theta_n)$}, depicted in 
    Fig.~\ref{fig_2dhisto}, has been obtained.
    Here and in the following figures error bars are shown only if the errors 
    are larger than the symbol size and dotted lines are guides to the eye.}
  \label{fig_clpdiag}
\end{figure}

At the critical field~$H_{\text{c}1}$, see Eq.~(\ref{eq_hc1}), the ground 
state is degenerate in the AF, the SF, and biconical structures, as 
illustrated in Fig.~\ref{fig_gstate}. This degeneracy in the biconical 
configurations, following from straightforward energy considerations, seems to 
have been overlooked in the previous analyses. The structures may be described 
by the tilt angles, $\theta_1$~and~$\theta_2$, formed between the directions 
of the spins on the two sublattices of the antiferromagnet and the easy axis. 
For a given value of~$\theta_1$, the other angle~$\theta_2$ is fixed by
\begin{equation}
  \theta_2 \; = \; \arccos 
  \left( \frac{ \sqrt{1-\Delta^2} \; - \; \cos\theta_1 }{ 1 \; - \; \sqrt{1-\Delta^2} \cos\theta_1 } \right) 
  \quad \text{.}
  \label{eq_bic}
\end{equation}
Obviously, the biconical configurations transform the AF into the SF 
state: The spins on the "up-sublattice" of the AF structure, with the
spins pointing into the direction of the field, may be thought of to vary 
from~$\theta_1=0$ to $\pm\theta_{\text{SF}}$, while the spins on the 
"down-sublattice" vary simultaneously from~$\theta_2=\pi$ 
to~$\mp\theta_{\text{SF}}$. Accordingly, $\theta_1$~determines 
uniquely~$\theta_2$ and vice versa. Apart from this continuous degeneracy 
in~$\theta_1$ (or~$\theta_2$), there is an additional rotational degeneracy of 
the biconical configurations in the spin components perpendicular to the easy 
axis, the $xy$-components, as for the SF structure, see 
Fig.~(\ref{fig_gstate}). These components are, of course, 
antiferromagnetically aligned at neighboring sites.
\begin{figure}
  \includegraphics{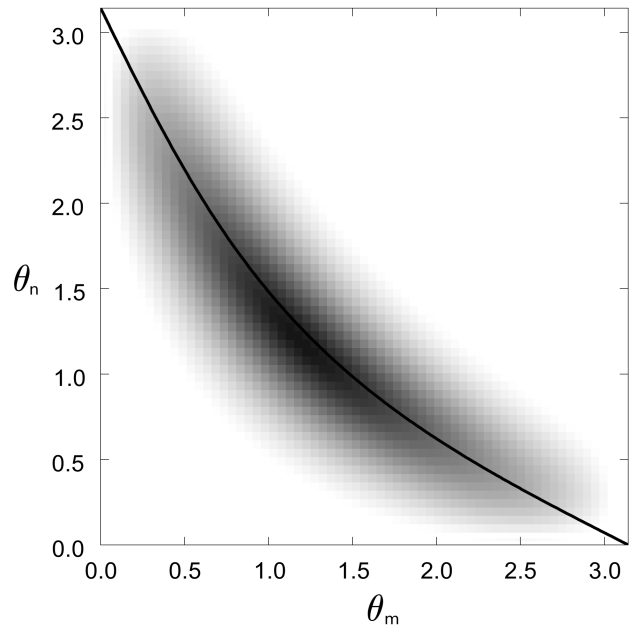}
  \caption{Joint probability distribution~\mbox{$P(\theta_m,\theta_n)$} 
    showing the correlations between the tilt angles~$\theta_m$ 
    and~$\theta_n$ on neighboring sites~$m$ and~$n$ for a system of 
    size~$L=80$ at~$H/J=2.41$, $k_BT/J=0.255$, and~$\Delta=\frac{4}{5}$. 
    \mbox{$P(\theta_m,\theta_n)$}~is proportional to the gray scale. The 
    superimposed black line depicts the relation between the two angles in the 
    biconical ground state, see Eq.~(\ref{eq_bic}).}
  \label{fig_2dhisto}
\end{figure}

To study the possible thermal relevance of the biconical structures at~$T>0$,
we performed Monte Carlo simulations analyzing the joint probability 
distribution~\mbox{$P(\theta_m,\theta_n)$} for having tilt 
angles~$\theta_m$ and~$\theta_n$ at neighboring sites, $m$~and~$n$. For 
comparison with the previous studies\cite{lb,ho,zl} we 
set~$\Delta=\frac{4}{5}$, leading to the phase diagram depicted in 
Fig.~\ref{fig_clpdiag}. For example, fixing the field at~$H=2.41J$, we 
observed at~\mbox{$k_BT/J\approx 0.255$} an Ising-type transition on approach 
from higher temperatures and a Kosterlitz-Thouless-type transition on approach 
from the low-temperature side, extending our corresponding previous 
findings\cite{ho} to even lower temperatures, and in agreement with recent 
results.\cite{zl} Indeed, as depicted in Fig.~\ref{fig_2dhisto}, in that part 
of the phase diagram, being in the vicinity of the very narrow intervening, 
supposedly disordered phase, the joint 
probability~\mbox{$P(\theta_m,\theta_n)$} exhibits a line of local maxima 
following closely Eq.~(\ref{eq_bic}), obtained for the ground state. That 
behavior is largely independent of the size of the lattices we 
studied. Similar signatures of the biconical structures are observed in the 
simulations at nearby temperatures, when fixing the field at~$H=2.41J$, as 
well as in the vicinity of the entire transition region between the AF and 
SF phases, see Fig.~\ref{fig_clpdiag}, at higher fields and temperatures.

Accordingly, we tend to conclude that biconical fluctuations are dominating in 
the narrow intervening phase. Whether that phase exists as a disordered phase 
down to the ground state or whether there is a stable biconical phase in two 
dimensions, remain open questions, being beyond the scope of this article.

Note that our additional Monte Carlo simulations for the anisotropic XY 
antiferromagnet in a field on a square lattice show that the analogues
of 'biconical' structures (the orientation of the spins being now given
by the two tilt angles only) and fluctuations play an important role
near the transition regime between the AF 
and SF phases in that case as well. In fact, Eq.~(\ref{eq_bic}) provides an 
excellent guidance for interpreting our simulational data similar to
the ones presented in Fig. 3.

\section{Quantum XXZ model}
\label{sec_qu}
\begin{figure}
  \includegraphics{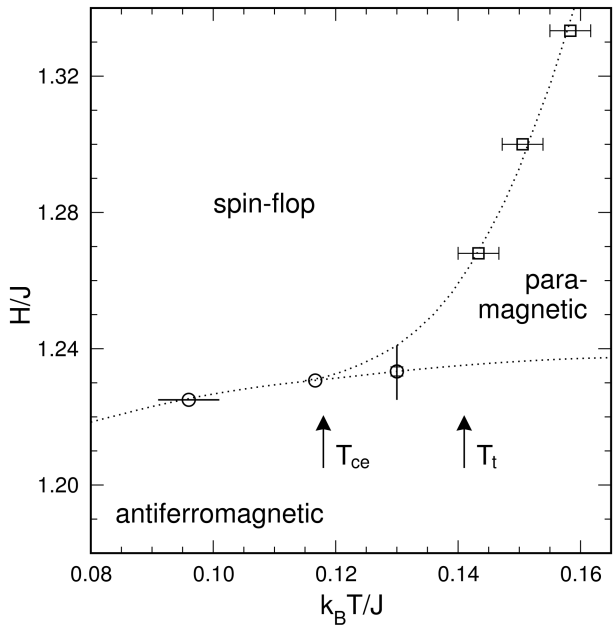}
  \caption{Phase diagram of the XXZ Heisenberg antiferromagnet with 
    spin-$\frac{1}{2}$ and~$\Delta=\frac{2}{3}$. The straight solid lines 
    denote the choices of parameters where our very extensive
    simulations, discussed in the text, have 
    been performed. The arrows mark the previously\cite{st} 
    suggested locations of the tricritical point~($T_{\text{t}}$) and 
    the critical endpoint~($T_{\text{ce}}$).}
  \label{fig_phdg}
\end{figure}

The aim of the study on the quantum version, $S=\frac{1}{2}$, of the XXZ 
model, Eq.~(\ref{eq_ham}), has been to check the previously suggested scenario 
of a first-order phase transition between the AF and SF phases extending up to 
a critical endpoint and with a tricritical point on the AF phase boundary, 
see Fig.~\ref{fig_phdg}.

We performed quantum Monte Carlo (QMC) simulations in the framework of the 
stochastic series expansion (SSE)\cite{sk} using directed loop 
updates~\cite{su}. We consider square lattices of $L\times L$~sites with the 
linear dimension~$L$ ranging from~$2$ to~$150$, employing full periodic 
boundary conditions. Defining, as usual,\cite{sk} a single QMC step as one 
diagonal update followed by the construction of several operator-loops, each 
individual run typically consists of~$10^6$ steps and is preceded by at 
least~$2\cdot10^5$ steps for thermal equilibration. Averages and error bars 
are obtained by taking into account results of several, ranging 
from~$8$ to~$32$, Monte Carlo runs, choosing different initial configurations 
and random numbers. Especially for large systems and low temperatures we 
additionally utilize the technique of parallel tempering (or exchange Monte 
Carlo)\cite{pt1,se} to enable the simulated systems to overcome the large 
energy barriers between configurations related to different phases more 
frequently. We typically work with a chain of $16$~to $32$~configurations in 
parallel which are simulated at different equally spaced temperature or 
magnetic field values allowing for an exchange of neighboring configurations 
after a constant number of QMC steps. The achieved reduction of the 
autocorrelation times, e.g. of the different magnetizations discussed
below, amounts up to several orders of magnitude and therefore results in 
significantly smaller correlations between subsequent measurements which, in 
turn, allows for shorter simulation times.

To determine the phase diagram and to check against previous work\cite{st}, we 
calculated various physical quantities including the $z$-component of the 
total magnetization,
\begin{equation}
  M^z \; = \; \frac{1}{L^2} \sum_i \langle S_i^z \rangle \quad \text{,}
\end{equation}
and the square of the $z$-component of the staggered magnetization,
\begin{equation}
  (M^z_{\text{st}})^2 \; = \; \frac{1}{L^2} 
  \left[ \sum_{i_a} \langle S_{i_a}^z \rangle \; - \; \sum_{i_b}
  \langle S_{i_b}^z \rangle \right]^2 
  \quad \text{,}
\end{equation}
summing over all sites, $i_a$ and $i_b$, of the two sublattices of the 
antiferromagnet. A useful quantity in studying the SF phase is the 
spin-stiffness~$\rho_s$ which is related to the change of the free-energy on 
imposing an infinitesimal twist on all bonds in one direction of the lattice. 
In QMC simulations the spin-stiffness can conveniently be measured by the 
fluctuations of the winding numbers~$W_x$ and~$W_y$,\cite{sk}
\begin{equation}
  \rho_s \; = \; \frac{k_B T}{2} \left( W_x^2 + W_y^2 \right) \quad \text{.}
\end{equation}
The winding numbers themselves are given by
\begin{equation}
  W_{\alpha} \; = \; \frac{1}{L} \left( N^+_{\alpha} \; - \; N^-_{\alpha} \right),
\end{equation}
where~$N^+_{\alpha}$ and~$N^-_{\alpha}$ denote the number of 
operators~$S^+_iS^-_j$ and~$S^-_iS^+_j$ in the SSE operator sequence with a 
bond~$\langle i,j \rangle$ in the $\alpha$-direction, $\alpha\in\{x,y\}$.

All data for the quantum model presented here are obtained at an anisotropy 
parameter of~$\Delta=\frac{2}{3}$ to allow for comparison with previous 
findings\cite{st,ho}. The phase diagram in the region of interest, where all 
three phases, the AF, the SF, and the paramagnetic phase 
occur, is displayed in Fig.~\ref{fig_phdg}.
\begin{figure}
  \includegraphics{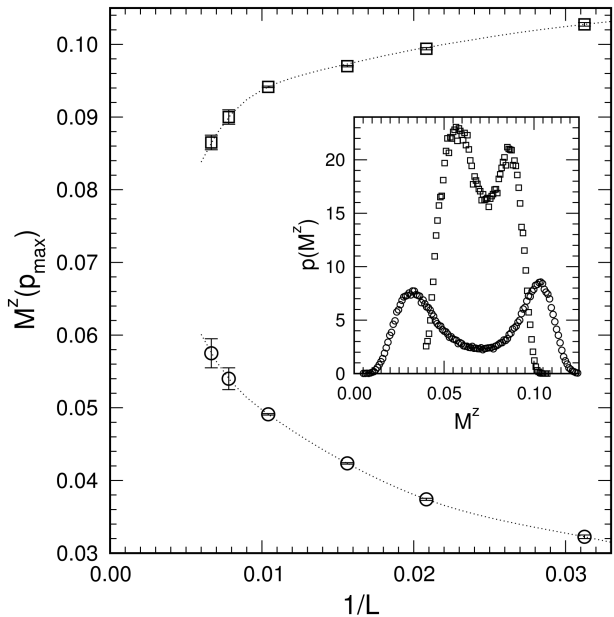}
  \caption{Positions of the maxima of the magnetization histograms as a 
    function of the inverse system size. The inset exemplifies two histograms 
    for systems of size $L=32$~(circles) and $L=150$~(squares) 
    at~$k_BT/J=0.13$ and the coexistence fields~$H/J=1.23075$ 
    and~$H/J=1.232245$.}
  \label{fig_hist}
\end{figure}

The earlier study\cite{st} asserted a phase diagram with a tricritical point 
at~\mbox{$k_B T_{\text{t}}/J\approx 0.141$} and a direct first-order 
transition between the SF and AF phases below the critical endpoint 
at~\mbox{$k_B T_{\text{ce}}/J\approx 0.118$}, see Fig.~\ref{fig_phdg}. In 
detail the authors identified a first-order AF to paramagnetic transition 
at~$k_BT/J=0.13$ by means of an analysis of the magnetization 
histograms~$p(M^z)$. We studied that case, improving the statistics and 
considering even larger lattice sizes. Indeed, as expected for a discontinuous 
change of the magnetization, the histograms of finite systems are confirmed to 
display two distinct maxima corresponding to the ordered and the disordered 
phase in the vicinity of the AF phase boundary (see inset of 
Fig.~\ref{fig_hist}). Note however, that such a two-peak structure can also be 
found for small systems at a continuous transition, with a single peak in the 
thermodynamic limit. Thence, a careful finite-size analysis is needed to, 
possibly, discriminate the two different scenarios. We simulated lattice sizes 
with up to~\mbox{$150\times 150$} spins adjusting the magnetic field such that 
coexistence of the phases, i.e. equal weight of the two peaks, is provided. As 
depicted in Fig.~\ref{fig_hist} the positions of the maxima as a function of 
the inverse system size exhibit a curvature, which becomes more pronounced for 
larger lattices. In contrast, in the previous analysis\cite{st} at the same 
temperature, linear dependences of the peak positions as a function of~$1/L$ 
had been presumed, leading to distinct two peaks in the thermodynamic limit. 
We conclude, that the previous claims of a first-order transition 
at~$k_BT/J=0.13$ and of the existence of a tricritical point 
at~\mbox{$k_BT_{\text{t}}/J\approx 0.141$} needs to be viewed with care. 
Indeed, the tricritical point seems, if it exists at all, to be shifted 
towards lower temperatures.

In the previous work\cite{st} a direct transition of first order 
between the AF and SF phases has been suggested to take place at 
lower temperatures, \mbox{$k_BT/J\leq k_BT_{\text{ce}}/J\approx 0.118$}. To 
check this suggestion we studied the system at constant field~$H/J = 1.225$, 
where such a direct transition would occur, see Fig.~\ref{fig_phdg}. 
Calculating the expectation values of the different magnetizations as well as 
the corresponding histograms we obtain an estimate of the critical temperature 
of the AF phase, \mbox{$k_BT_{\text{AF}}=0.09625\pm 0.0005$}.
\begin{figure}
  \includegraphics{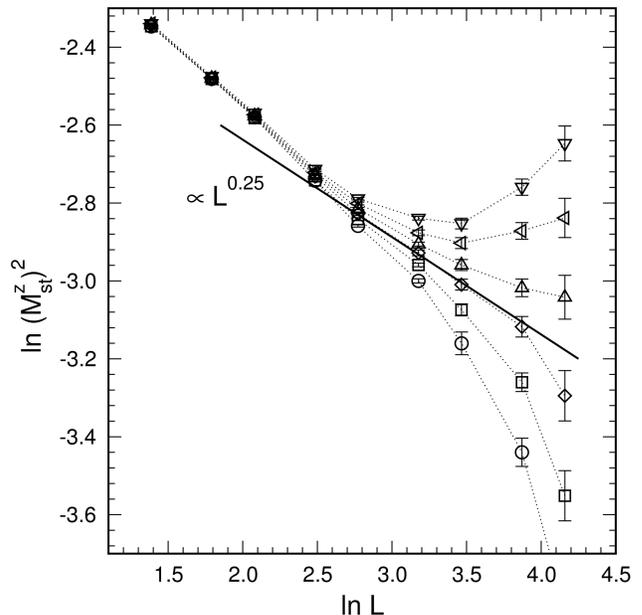}
  \caption{Doubly logarithmic plot of the staggered 
    magnetization~$(M^z_{st})^2$ vs. the system size~$L$ at~$H/J=1.225$ for 
    the temperatures $k_BT/J = 0.095$, $0.0955$, $0.096$, $0.0965$, $0.097$, 
    and~$0.0975$ (from bottom to top). The straight line proportional 
    to~$L^{\frac{1}{4}}$ illustrates the expected finite-size behavior close 
    to a continuous transition of Ising type.}
  \label{fig_stmz}
\end{figure}

Surprisingly, approaching the transition from the AF phase, the finite-size 
behavior of the squared staggered magnetization~$(M^z_{\text{st}})^2$, 
being the AF order parameter, is still consistent with a continuous transition 
in the Ising universality class: As illustrated in Fig.~\ref{fig_stmz} the 
asymptotic region is very narrow, similar to the observations in the classical 
model.\cite{ho,zl} The dependence on the system size seems to 
obey~\mbox{$(M^z_{\text{st}})^2 \propto L^{1/4}$} right at the transition, as 
expected for the Ising universality class.\cite{on}

Furthermore, approaching the transition from the SF phase, an analysis 
of the spin-stiffness~$\rho_s$  at 
the same field value of~$H/J=1.225$ results in about the same transition 
temperature, \mbox{$k_BT_{\text{SF}}/J=0.09625\pm 0.001$}. Thence, there may 
be either a unique transition between the SF and AF phases, or, as observed 
in the classical case, an extremely narrow intervening phase, with phase 
boundaries of Ising and Kosterlitz-Thouless (KT) type.

\begin{figure}
  \includegraphics{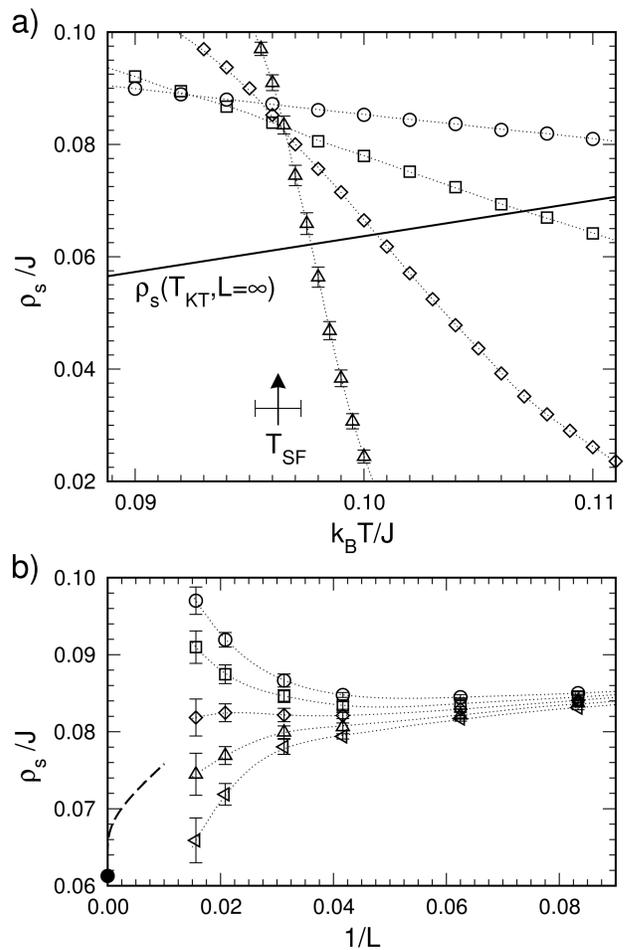}
  \caption{a) Spin stiffness~$\rho_s/J$ vs. temperature~$k_BT/J$ for the 
    different system sizes $L=8$~(circles), $16$~(squares), $32$~(diamonds), 
    and $64$~(triangles). The straight line denotes the critical value of the 
    spin-stiffness according to the formula of Nelson and Kosterlitz\cite{nk}, 
    see Eq.~(\ref{eq_ktstif}).\\
    b) Finite-size behavior of the spin-stiffness~$\rho_s/J$ at $H/J=1.225$ as 
    a function of the inverse system size,~$1/L$ for the 
    temperatures~$k_BT/J=0.0955$, $0.096$, $0.0965$, $0.097$, and~$0.0975$ 
    (from top to bottom). The dashed curve illustrates the estimated 
    asymptotic behavior according to Eq.~(\ref{eq_stifffs}) 
    with~$k_BT_{\text{KT}}/J=0.09625$ and~$C_0=5$, the corresponding critical 
    value~\mbox{$\rho_s(T_{\text{KT}},L=\infty)\approx 0.0613$} is marked by 
    the filled circle.}
  \label{fig_stif}
\end{figure}

To determine, whether a KT transition describes the disordering of the SF 
phase, we check the theoretical prediction\cite{kt,nk} that for the infinite 
system the spin-stiffness is finite within the SF phase, 
takes on a universal value at the KT transition related to~$T_{\text{KT}}$ by
\begin{equation}
  \rho_s(T=T_{\text{KT}},L=\infty) \; = \; \frac{2}{\pi} \; k_B T_{\text{KT}} \quad \text{,}
  \label{eq_ktstif}
\end{equation}
and discontinuously vanishes in the disordered phase. As depicted in 
Fig.~\ref{fig_stif}, the spin-stiffness~$\rho_s$ at~$T = T_{\text{SF}}$ seems 
to be, at first sight, significantly larger than the KT-critical value given 
by Eq.~(\ref{eq_ktstif}). Indeed, in the earlier study\cite{st} it has been 
argued, based on similar observations, that there is a direct first order AF to SF 
transition. However, the finite-size effects close to the 
transition deserve a careful analysis: For the KT scenario, renormalization 
group calculations\cite{wm,wm1} predict the asymptotic size dependence 
at~$T=T_{\text{KT}}$ to obey 
\begin{multline}
  \rho_s(T=T_{\text{KT}},L) \quad = \\
  \quad \rho_s(T=T_{\text{KT}},L=\infty) \; \left( 1 \; + \; \frac{1}{2\ln L \; - \; C_0} \right) \quad \text{,}
  \label{eq_stifffs}
\end{multline}
where~$C_0$ denotes an apriorily unknown, non-universal, parameter. By 
studying the quantity\cite{hk}
\begin{equation}
  C(L) \; = \; -2 \left[ \left( \frac{\pi \rho_s}{k_B T} - 2 \right)^{-1} \; - \; \ln L \,\right] \quad \text{,}
  \label{eq_stifcl}
\end{equation}
which, according to Eqs.~\ref{eq_ktstif} and~\ref{eq_stifffs}, converges 
for~$L\rightarrow\infty$ and~$T=T_{\text{KT}}$ to the value~$C_0$ at a KT 
transition, we obtain a rough estimate of $C_0 \approx 5$. A prediction of the 
finite-size behavior at~$T_{\text{SF}}$ is obtained by inserting this value, 
$C_0=5$, into Eq.~(\ref{eq_stifffs}). Comparing the data of the 
spin-stiffness~$\rho_s$ in the direct vicinity of the boundary of the SF phase 
with the prediction according to the KT theory, see Fig.~\ref{fig_stif} b), 
one may conclude that the lattice sizes accessible by simulations, 
\mbox{$L\leq 64$}, seem to be too small to capture the asymptotic finite-size 
behavior. In any event, in case of a KT transition, the 
spin-stiffness~$\rho_s$ drops asymptotically very rapidly to its universal 
critical value as a function of system size, being consistent with the 
relatively large values for the simulated finite lattices. Thus, a scenario 
with a KT transition between the SF and a narrow intervening disordered phase 
cannot be ruled out by the present large-scale simulations down to 
temperatures as low as\mbox{~$k_BT/J=0.09625 \pm 0.001$}.

Of course, it is desirable to quantify the role of biconical fluctuations in 
the quantum case as well. However, accessing the probability distributions of 
the tilt angles studied in Sect.~\ref{sec_cl} for the quantum case is beyond 
the scope of the present numerical analysis.

\section{Summary}
\label{sec_ds}

We studied the classical and quantum, $S=\frac{1}{2}$, versions of the XXZ 
Heisenberg antiferromagnet on the square lattice in an external field along 
the easy axis. The model is known to display ordered AF and SF as well as 
disordered, paramagnetic phases. Here we focused attention to the region of 
the phase diagram near and below the temperature where the two boundary lines 
between the AF and the SF phases and the disordered phase approach 
each other, meeting eventually at a triple point. We performed Monte Carlo 
simulations, augmented, in the classical case, by a ground state analysis. 

In the classical version, we presented first direct evidence for the 
importance of biconical structures in the XXZ model. Indeed, such 
configurations do exist already as ground states at the critical 
field~$H_{\text{c}1}$, separating the AF and SF phases. The interdependence of 
the two tilt angles, characterizing the biconical ground states, persists
at finite temperatures, in the region where the narrow phase between the AF 
and SF phases is expected to occur. Indeed, the joint probability distribution 
of the tilt angles at neighboring sites demonstrates the thermal significance 
of those configurations. Previous arguments on $O(3)$~symmetry in that 
region and down to zero temperature thus have to be viewed with care.
The results of the present simulations suggest that, if the biconical 
configurations do not lead to a stable biconical phase in two dimensions, the 
narrow intervening phase is a disordered phase characterized by biconical 
fluctuations. In this sense the "hidden bicritical point" at~$T=0$ may then be 
coined into a "hidden tetracritical point."

In the quantum version, previous simulations suggested, on lowering the 
temperature, the existence of a tricritical point on the boundary line between 
the AF and disordered phases, followed by a critical endpoint being the triple 
point of the AF, SF and disordered phases, and eventually by a first-order 
transition between the AF and SF phases at sufficiently low temperatures. The 
present simulations, considering larger system sizes and improved statistics, 
provide evidence that this scenario, if it exists at all, has to be shifted 
to lower temperatures than proposed before. Of course, simulations on even 
larger lattices and lower temperatures would be desirable, but are extremely 
time consuming. 

A clue on possible distinct phase diagrams for the classical and quantum 
versions may be obtained from an analysis of biconical fluctuations in the 
quantum case. Experimental studies on signatures of those fluctuations are 
also encouraged.

\acknowledgments
Financial support by the Deutsche Forschungsgemeinschaft under grant 
No.~SE~324/4 is gratefully acknowledged. We thank A.~Honecker,
B.~Kastening, R.~Leidl, A.~Pelissetto, M.~Troyer, and E.~Vicari for
useful discussions and 
information.

\end{document}